%% file: Manuscript.tex
\documentclass[%
reprint,
 amsmath,amssymb,
 aps,
]{revtex4-1}

\usepackage{braket}
\usepackage{siunitx}
\usepackage{nicefrac}
\usepackage{graphicx}
\usepackage{dcolumn}
\usepackage{bm}
\usepackage{xcolor}
\usepackage[makeroom]{cancel}
\usepackage{mathtools}
\usepackage{comment}

\usepackage{enumitem}
\usepackage{interval}
\DeclareSIUnit\barn{b}

\begin{document}

\preprint{APS/123-QED}

\title{Nuclear Excitation by Electron Capture in Excited Ions}

\author{Simone Gargiulo}%
\email{simone.gargiulo@epfl.ch}

\author{Ivan Madan}%
\author{Fabrizio Carbone}%

\email{fabrizio.carbone@epfl.ch}
\affiliation{%
 Institute of Physics (IPhys), Laboratory for Ultrafast Microscopy and Electron Scattering (LUMES),\\
École Polytechnique Fédérale de Lausanne (EPFL), Lausanne 1015 CH, Switzerland
}%


\begin{abstract}
A nuclear excitation following the capture of an electron in an empty orbital has been recently observed for the first time. So far, the evaluation of the cross section of the process has been carried out widely using the assumption that the ion is in its electronic ground state prior to the capture. We show that by lifting this restriction new capture channels emerge resulting in a boost of more than three orders of magnitude to the electron capture resonance strength.
\end{abstract}

\maketitle


Innovative technologies for harvesting and long-duration storing of energy are currently highly desired \cite{Koningstein2014,sepulveda2021design}. In this context,
isomers are particularly attractive as they provide the potential for on-demand clean energy release combined with reliability, compactness, high stored energy density and the ability to operate in extreme environment.
The achievement of a controlled and efficient extraction of the isomeric energy has been a milestone for decades and is recently attracting growing attention \cite{Walker1999a,collins1999accelerated,litz2004controlled,collins2004accelerated,aprahamian2005long,dzyublik2013triggering,carroll2013nuclear,vanacore2018attosecond}.
In particular, recently demonstrated Nuclear Excitation by Electron Capture (NEEC) \cite{Chiara2018} could possibly offer gains in terms of control, as the electron switch of the process can be manipulated by means of electron optics and wave function engineering \cite{vanacore2018attosecond,Madan2020}.

NEEC is a process in which the capture of a free or target electron by an ion results in the resonant excitation of a nucleus.
The kinetic energy of the free electron, $E_\mathrm{r}$, needs to equal the difference between the nuclear transition energy, $E_\mathrm{n}$, and the atomic binding energy released through electron capture, $E_\mathrm{b}$ (i.e., $E_\mathrm{r}=E_\mathrm{n} - E_\mathrm{b}$). The first isomer depletion induced by electron capture was recorded in a beam-based setup in 2018 \cite{Chiara2018}, albeit the strength of the detected signal is unexplained by state-of-the-art theory \cite{Wu2019}, presenting a discrepancy of about nine orders of magnitude.
Till today, NEEC is  an object of a live debate \cite{guo2021possible,chiara2021reply,rzadkiewicz2021novel}.

Until this work, the NEEC process has been considered only in ions which are in their electronic ground states (ground state assumption, GSA) \cite{Palffy2008,wu2018tailoring,Gunst2018,Rzadkiewicz2019}, in ground state ions with a single inner-shell hole created by X-rays \cite{wu2019x} or considering a statistical approach for electronic populations in an average atom model \cite{gosselin2004enhanced,doolen1978nuclear,doolen1978nuclearC}.
In this Letter we examine the role of excited electronic configurations without any restrictions on the initial levels population.
While the GSA allows for a straightforward account of the capture channels, it is too restrictive to unequivocally represent the real conditions taking place in out of equilibrium scenarios.
In fact, it has been shown that, for a given charge state $q$, the ground state configuration usually is not the most probable \cite{son2014quantum}. It is therefore important to evaluate the cross sections of nuclear processes for a wider range of electronic configurations.

The GSA rules out the capture in the innermost shells for partially filled ions. For example, one can have K-capture till two electrons fill the 1s orbital.
However, even for fully ionized nuclei, NEEC into K-shell may be forbidden if the energy released through a K-capture ($E_\mathrm{b}^\mathrm{K}$) exceeds the nuclear transition energy (i.e., $E_\mathrm{r}<0$). Therefore, for such nuclei, under the GSA, NEEC with capture in the K-shell is never possible.
These channels can be re-enabled if sufficient screening is provided by an out of equilibrium electronic configuration, as we show for the example of $^{73}$Ge.

\begin{figure}[!htbp]
\includegraphics[width=\linewidth]{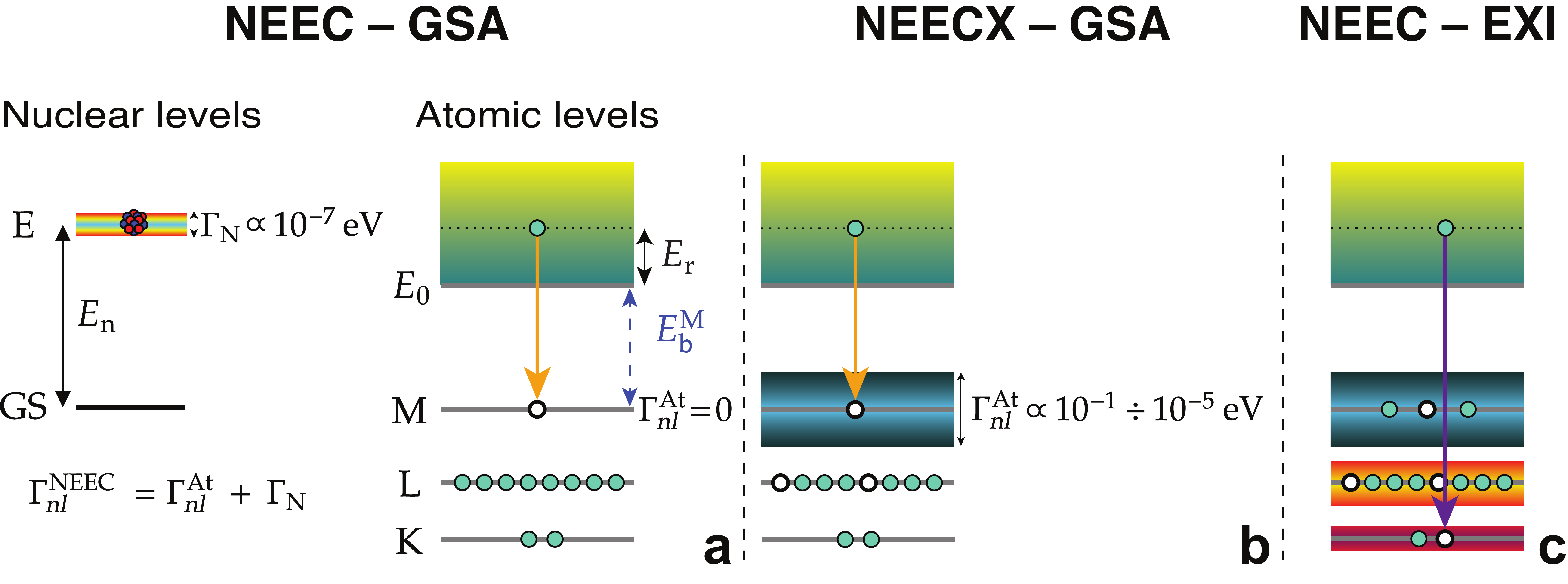}
\caption{\label{fig:NEEC_Config} Atomic configurations in case of electron capture: conventionally, the ion is considered to be in its nuclear and electronic ground states, while the capture either leaves the ion in the electronic ground state (a) referred to as NEEC, or bring it in an electronic excited state (b), referred to as NEECX.
In (c), electrons can be distributed all over K, L and M shells. $\Gamma$ represents the width of the atomic ($\Gamma_{nl}^\mathrm{At}$) and nuclear ($\Gamma_\mathrm{N}$) transitions. For atomic ground states $\Gamma_{nl}^\mathrm{At}=0$, while for excited configurations $\Gamma_{nl}^\mathrm{At}\gg \Gamma_\mathrm{N}$.}
\end{figure}

In Fig. \ref{fig:NEEC_Config} we compare both the conventional and our approach.
In Fig. \ref{fig:NEEC_Config}a NEEC takes place in an ion under the GSA. A variant of NEEC --- i.e., NEEC followed by a fast x-ray emission (NEECX) --- considers the capture of the electron in a higher energy electronic shell while the ion is still in its electronic ground state, a situation in which the GSA still holds \cite{Palffy2008,Polasik2017}, see Fig. \ref{fig:NEEC_Config}b.
Instead, Fig. \ref{fig:NEEC_Config}c represents the case in which the GSA does not hold: here, NEEC can occur even in excited ions (NEEC-EXI) and the consequences of such a scenario are discussed below.

\noindent

The integrated NEEC cross section, called resonance strength $S_\mathrm{NEEC}$, can be expressed as \cite{wong2008introductory, Zadernovsky2002, Harston,KineticPhotons,Palffy2006,Gunst2018}:

\begin{equation}
 S_{\mathrm{NEEC}}^{q,\alpha_\mathrm{r}}  = \int \mathrm{d}E\ \dfrac{\lambda_\mathrm{e}^2}{2}\ \hbar\ Y_{\mathrm{NEEC}}^{q,\alpha_\mathrm{r}}(E)\, L_\mathrm{r}(E-E_\mathrm{r}) ,
\label{eq:NEEC-GSA}
\end{equation}
\noindent
where $\lambda_\mathrm{e}$ is the electron wavelength and $L_\mathrm{r}$ is a Lorentzian function centered at the resonance energy of the free electron $E_\mathrm{r}$. The width of such Lorentzian is given by the combination of both atomic configuration and nuclear level, $\Gamma^\mathrm{NEEC}_{nl}=\Gamma^\mathrm{At}_{nl}+\Gamma_\mathrm{N}$. $Y_{\mathrm{NEEC}}^{q,\alpha_\mathrm{r}}$ is the microscopic NEEC rate  that depends on the final electronic configuration ($\alpha_\mathrm{r}$) and on the ion charge state $q$ prior to the electron capture.
Under the GSA, the initial electronic configuration ($\alpha_0)$ is uniquely defined by the charge state $q$ and the number of available channels for capture in a particular subshell $nl_j$ is strongly limited.
By contrast, in NEEC-EXI the rate $Y_{\mathrm{NEEC}}^{q,\alpha_\mathrm{r}}$ also depends on $\alpha_0$, thus it has to be expressed as $Y_{\mathrm{NEEC}}^{q,\alpha_0,\alpha_\mathrm{r}}$.

In NEEC-EXI, for a given charge state $q$, electrons are assigned to a particular shell from the innermost to the outermost (K, L, M) encompassing all possible combinations.
All these states are used as initial configurations $\alpha_0$.
In case the electron involved in the capture breaks the orbital angular momentum coupling in the initial atomic configuration $\alpha_0$, the expression of the NEEC resonance strength in Eq. \ref{eq:NEEC-GSA} is further complicated by an additional coefficient $\Lambda$, expressing the recoupling probability between the initial ($\alpha_0$) and final electronic configurations ($\alpha_\mathrm{r}$) \cite{Bilous2017,yutsis1962mathematical,biedenharn1981racah,Aquilanti2008,Fack1995,Fack1999}:

\begin{equation}
 S_{\mathrm{NEEC}}^{q,\alpha_\mathrm{0},\alpha_\mathrm{r}} = \Lambda\ \int \mathrm{d}E\ \dfrac{\lambda_\mathrm{e}^2}{2}\ \hbar \ Y_{\mathrm{NEEC}}^{q,\alpha_\mathrm{0},\alpha_\mathrm{r}}(E)\, L_\mathrm{r}(E-E_\mathrm{r}).
\label{eq:NEEC-EXI}
\end{equation}
\noindent
In this Letter, the recoupling schemes for ions with up to four electrons filling the orbitals have been considered.
Further details about the expression of $\Lambda$ and electron recoupling are given in the Supplemental Material \cite{SM}.
The microscopic NEEC rate $Y_\mathrm{NEEC}$ is related to the process of internal conversion by time reversal. Using the principle of detailed balance \cite{KineticPhotons}, $Y_\mathrm{NEEC}$ can be expressed as a function of the internal conversion coefficients (ICCs) $\alpha_\mathrm{IC}$.

The determination of the ICCs for ions requires the knowledge of the electronic configuration and of the bound and free electron wavefunctions. In first approximation, ICCs for ions can be estimated from those of neutral atoms applying a scaling procedure, which relates ICC with the binding energy and occupancy of a specific subshell \cite{Rzadkiewicz2019,rysavy2000reliability,gosselin2004enhanced,larkins1977semiempirical,sevier1979atomic,kantele1989simplified}. In this case, ICCs for neutral atoms are theoretically computed using the frozen orbital (FO) approximation based on the Dirac-Fock calculations \cite{kibedi2008evaluation}. Albeit ICCs for neutral atoms have been shown to have less than $1\%$ uncertainty compared to experimental data \cite{kibedi2007internal,raman2002good,kibedi2008evaluation}, no detailed uncertainty analysis has been performed on ions for this scaling procedure.

\begin{figure*}[ht]
\includegraphics[width=\linewidth]{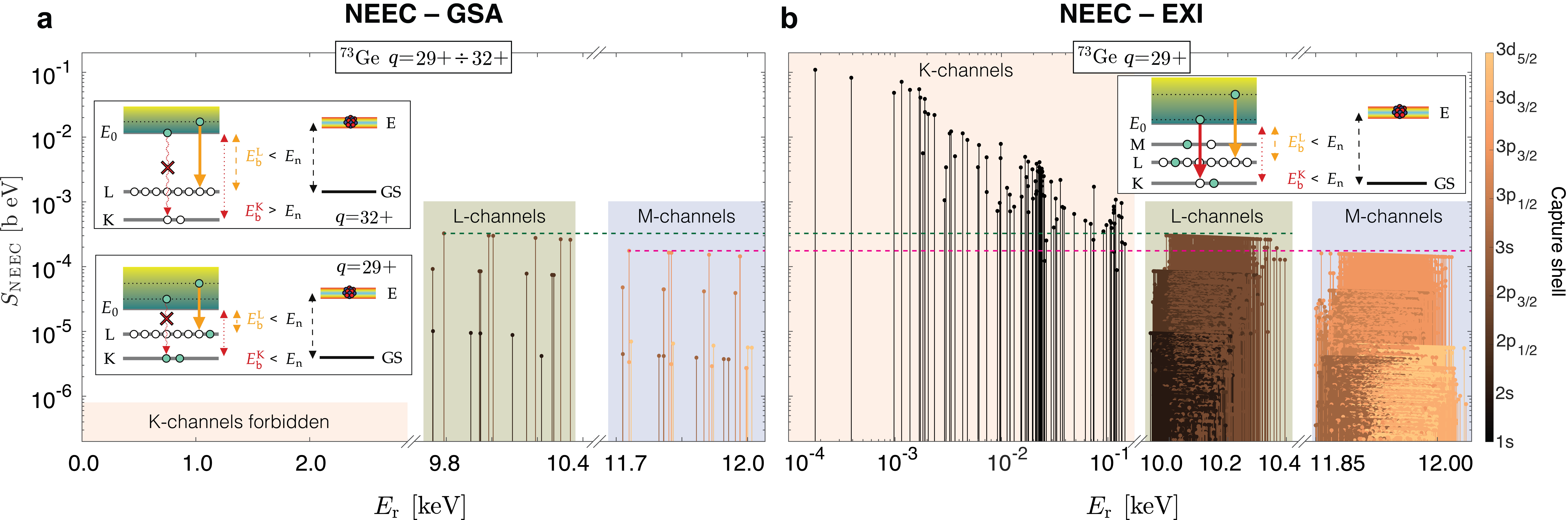}   
\caption{\label{fig:NEEC_Cross_Sections} (a) Resonance strengths for capture in the L- (green box) and M-shell (blue box) in case of $^{73}$Ge with $q=\interval{29+}{32+}$. NEEC in the K-shell is energetically forbidden (pink box), since for high charge states the binding energy released for a K-capture is bigger than the nuclear transition ($E_\mathrm{b}^\mathrm{K}>E_\mathrm{n}$). For $q=$ 29+ $E_\mathrm{b}^\mathrm{K}<E_\mathrm{n}$, however the K-shell is completely filled and capture can not occur (insets). (b) Resonance strengths for $^{73}$Ge in case all the possible combinations of initial and final electronic configurations are taken into account, for $q=$ 29+.
Each resonant channel is represented by a solid line, with its colors indicating the capture orbital.
The horizontal green and magenta lines indicate the highest $S_\mathrm{NEEC}$, under GSA, for L- and M-shells --- occurring at $[q,\alpha_\mathrm{r}]=\interval{32+}{2\mathrm{p}_{3/2}}$ and $[q,\alpha_\mathrm{r}]=\interval{32+}{3\mathrm{p}_{3/2}}$ --- respectively.}
\end{figure*}
\noindent

For this reason, we compute $Y_\mathrm{NEEC}$ of selected channels also with the more advanced theory presented in Ref. \onlinecite{Gagyi-Palffy06}, based on Feshbach projection operator formalism and compare these results with the ones obtained from the ICCs scaling procedure. Binding energies for a specific subshell and wavefunctions for a specific atomic configuration are computed using FAC \cite{gu2008flexible}. FAC is a fully relativistic atomic code taking configuration interaction into account. Accuracy for the computed energy levels is assessed to be in the order of few eV \cite{massacrier2012extensive}.

Applying the GSA to the $^{73}$Ge nuclear transition of $E_\mathrm{n}=\SI{13.2845}{\kilo\electronvolt}$ between the $9/2^+$ ground and the $5/2^+$ first excited states provides 47 L- and M-channels for $q=\interval{29+}{32+}$, shown in Fig. \ref{fig:NEEC_Cross_Sections}a and tabulated in the Supplemental Material \cite{SM}. Here, the K-shell is energetically forbidden and L-channels are the innermost available. The GSA allows for a drastic reduction of computational effort, as by lifting it, a total of 32723 capture channels can be found in the same charge state range for L- and M-shells.
Moreover, upon filling the orbitals, the electron screening lowers the binding energy of the K-shell. Once $E_\mathrm{b}^\mathrm{K}$ becomes smaller than $E_\mathrm{n}$, NEEC into the K shell is possible. For $^{73}$Ge this condition is met for $q = 29+$,  for which $100$ K-capture channels have been unveiled, as shown in Fig. \ref{fig:NEEC_Cross_Sections}b. Most of these K-channels (78) are characterized by an initial electronic configuration $\alpha_0$ of the type $1\mathrm{s}^1\ 2nl_j^1\ 3nl_j^1$ and occur in the energy range $E_\mathrm{r}= [\SIrange[range-phrase ={, } ,range-units = single]{0}{38.8}{]\ \electronvolt}$, while for the remaining ones $\alpha_0$ is $1\mathrm{s}^1\ 2nl_j^2$ and $E_\mathrm{r}= [\SIrange[range-phrase ={, } ,range-units = single]{48.2}{144.6}{]\ \electronvolt}$. All these K-channels have one electron in the K-shell prior to the electron capture: in fact, K-capture with $\alpha_0 = \{2nl_j^3\}$ are still forbidden, since $E_\mathrm{b}^\mathrm{K}$ is larger than $E_\mathrm{n}$ by about \SI{200}{\electronvolt} at $q=29+$. Resonance strengths for higher charge states are shown in the Supplemental Material \cite{SM}. For L and M channels the widths of the atomic configurations $\Gamma_\mathrm{nl}^\mathrm{At}$ is much smaller than the resonance energy $E_\mathrm{r}$ and $Y_\mathrm{NEEC}$ can be moved out of the integral in Eq. \ref{eq:NEEC-EXI}. In the case of K-channels instead $\Gamma_\mathrm{nl}^\mathrm{At}\approx E_\mathrm{r}$ and the integral of $Y_\mathrm{NEEC}$ has to be performed. The widths for the atomic configurations leading to a K-capture have been calculated using the XATOM code \cite{jurek2016xmdyn, son2011impact, murphy2014femtosecond}. Notably, the higher number of channels identified in NEEC-EXI is not only due to the several initial configurations considered, but also to the increase of the capture channels available for a single excited configuration $\alpha_0$ compared to the ground state counterpart. The reason is that excited configurations can have a larger number of open shells, thus the number of final configurations that can be generated are generally more numerous due to the higher number of combinations possible for the electron couplings.

Fig. \ref{fig:NEEC_Cross_Sections} compares the resonance strengths of the newly opened K-channels with the L- and M-channels for NEEC-GSA and NEEC-EXI.
Here, only shells up to the M have been considered, since $\alpha_{0}$ with electrons in higher shells do not provide sufficient screening for a K-capture at $q=29+$.
Selected channels are reported in Table \ref{tab:NEEC_Generic} and, when possible, are compared with those evaluated with the GSA procedure for which Eq. \ref{eq:NEEC-EXI} reduces to Eq. \ref{eq:NEEC-GSA}, since $\Lambda=1$, and results coincide.
Results for NEEC-EXI have been also evaluated using the wavefunction formalism (indicated as WF) and reported in Table \ref{tab:NEEC_Generic}. The maximum value obtained in this case is of \SI{5.18e-1}{\barn \electronvolt}. This allows us to comment on the accuracy of the ICC scaling procedure. Although the resonance strength obtained for a specific channel can be inaccurate by one order of magnitude, the ICC scaling reproduces the overall trend and the higher cross-section values in the case of NEEC-EXI.
Furthermore, it allows to have an easy estimate of the order of magnitude of the NEEC cross-section in different experimental scenarios.

\begin{table*}[btp]
\caption{\label{tab:NEEC_Generic} Resonance strengths for $^{73}$Ge in case of NEEC-EXI considering both the ICCs scaling procedure and the wavefunction (WF) formalism. For a given final electronic state ($\alpha_r$) all the relative parent configurations $\alpha_0$, that through electron capture can lead to it, are taken into consideration. Resonance energies are intended as $E_\mathrm{r}=E_\mathrm{n}-E_\mathrm{b}$. When possible, a comparison with the conventional derivation (GSA) is also presented. In blue the subshell $nl_j$ in which the capture occurs.}
\begin{ruledtabular}
\begin{tabular}{ccccc|cc}
 \multicolumn{4}{c}{$^{73}$Ge}& NEEC-GSA & \multicolumn{2}{c}{NEEC-EXI}\\
 $q$ &Initial Configuration ($\alpha_0$) &Final Configuration ($\alpha_\mathrm{r}$) & $E_\mathrm{r}$ [eV] &$S_\mathrm{NEEC}$ [b$\cdot$eV] & \multicolumn{2}{c}{$S_\mathrm{NEEC}$ [b$\cdot$eV]}\\[0.7mm]
\multicolumn{4}{c}{}& {ICCs scaling} & {ICCs scaling} & {WF}
\\ \hline\\[0.05mm]
 32+&$-$&$\color{blue}2\mathrm{p}_{3/2}^1\color{black}$ & $\SI{9.79e3}{}$ & \SI{3.26e-04}{} & \SI{3.26e-04}{} & \SI{1.25e-04}{}\\[2mm]
  32+&$-$&$\color{blue}2\mathrm{p}_{1/2}^1\color{black}$ & $\SI{9.74e3}{}$ & \SI{9.21e-05}{} & \SI{9.21e-05}{} & \SI{6.76e-05}{}\\[2mm]
 31+&$1\mathrm{s}^1$ & $1\mathrm{s}^1\ \color{blue}2\mathrm{s}^1 \color{black}$ &$\SI{9.91e3}{}$& \SI{9.49e-6}{}&  \SI{9.49e-6}{} &  \SI{8.37e-06}{}\\[2mm] 
 31+&$3\mathrm{d}_{3/2}^1$ &\color{blue}$2\mathrm{p}_{1/2}^1\color{black}\ 3\mathrm{d}_{3/2}^1$ &$\SI{9.82e3}{}$ & Not Allowed& \SI{8.92e-5}{} & \SI{6.56e-05}{}\\[2mm] 
 30+&$1\mathrm{s}^2$&$1\mathrm{s}^2\ \color{blue} \ 2\mathrm{s}^1 \color{black}$ & \SI{10.09e3}{} & \SI{8.82e-06}{} &\SI{8.82e-06}{} &\SI{7.47e-06}{}\\[2mm] 
   30+&$3\mathrm{d}_{3/2}^1\ 3\mathrm{d}_{5/2}^1$ & $\color{blue} 2\mathrm{p}_{3/2}^1 \color{black} \ 3\mathrm{d}_{3/2}^1\ 3\mathrm{d}_{5/2}^1$ & \SI{9.96e3}{} & Not Allowed &\SI{2.67e-04}{} &\SI{1.05e-04}{}\\[2mm] 
 29+&$1\mathrm{s}^2\ 2\mathrm{s}^1$&$1\mathrm{s}^2\ 2\mathrm{s}^1 \ \color{blue}2\mathrm{p}_{1/2}^1\color{black}$ & \SI{10.26e3}{} & \SI{7.46e-5}{} &  \SI{7.46e-5}{}  & \SI{5.19e-05}{}\\[2mm] 
 29+&$3\mathrm{p}_{3/2}^1\ 3\mathrm{d}_{3/2}^1 \ 3\mathrm{d}_{5/2}^1 $ & $\color{blue}2\mathrm{p}_{3/2}^1\color{black}\ 3\mathrm{p}_{3/2}^1\ 3\mathrm{d}_{3/2}^1 \ 3\mathrm{d}_{5/2}^1 $  & \SI{10.06e3}{} & Not Allowed &  \SI{6.30e-5}{}  &  \SI{2.53e-05}{}\\[2mm] 
  29+&$3\mathrm{p}_{3/2}^1\ 3\mathrm{d}_{3/2}^1 \ 3\mathrm{d}_{5/2}^1 $ & $\color{blue}2\mathrm{p}_{1/2}^1\color{black}\ 3\mathrm{p}_{3/2}^1\ 3\mathrm{d}_{3/2}^1 \ 3\mathrm{d}_{5/2}^1 $ & \SI{10.03e3}{} & Not Allowed &  \SI{7.18e-5}{} &  \SI{ 5.58e-05}{}\\[2mm] 
 29+&$1\mathrm{s}^1\ 2\mathrm{p}_{1/2}^1\ 2\mathrm{p}_{3/2}^1$&$\color{blue}1\mathrm{s}^2\color{black}\ 2\mathrm{p}_{1/2}^1\ 2\mathrm{p}_{3/2}^1$  & \SI{144.62}{} & Not Allowed &  \SI{2.23e-4}{}&{\SI{9.40e-04}{}}\\[2mm] 
 29+&$1\mathrm{s}^1\ 2\mathrm{s}^2$&$\color{blue}1\mathrm{s}^2\color{black}\ 2\mathrm{s}^2$ & \SI{73.87}{} & Not Allowed &  \SI{1.71e-3}{}&{\SI{7.28e-03}{}}\\[2mm] 
 29+&$1\mathrm{s}^1\ 2\mathrm{s}^1\  3\mathrm{d}_{5/2}^1$&$\color{blue}1\mathrm{s}^2\color{black}\ 2\mathrm{s}^1\ 3\mathrm{d}_{5/2}^1$  &  \SI{4.32}{}& Not Allowed &  \SI{1.14e-2}{}&{\SI{4.94e-2}{}}\\[2mm] 
 29+&$1\mathrm{s}^1\ 2\mathrm{s}^1\ 3\mathrm{p}_{1/2}^1$&$\color{blue}1\mathrm{s}^2\color{black}\ 2\mathrm{s}^1\ 3\mathrm{p}_{1/2}^1$  & \SI{1.15}{} & Not Allowed &\SI{7.14e-2}{}&{\SI{3.10e-1}{}}\\[2mm] 
  29+&$1\mathrm{s}^1\ 2\mathrm{p}_{3/2}^1\ 3\mathrm{d}_{3/2}^1$&$\color{blue}1\mathrm{s}^2\color{black}\ 2\mathrm{p}_{3/2}^1\ 3\mathrm{d}_{3/2}^1$  &  \SI{0.98}{}& Not Allowed &  \SI{4.82e-2}{}&{\SI{2.32e-1}{}}\\[2mm ]
 29+&$1\mathrm{s}^1\ 2\mathrm{p}_{3/2}^1\ 3\mathrm{d}_{3/2}^1$&$\color{blue}1\mathrm{s}^2\color{black}\ 2\mathrm{p}_{3/2}^1\ 3\mathrm{d}_{3/2}^1$  &  \SI{0.39}{}& Not Allowed &  \SI{8.25e-2}{}&{\SI{3.61e-1}{}}\\[2mm ]
 29+&$1\mathrm{s}^1\ 2\mathrm{s}^1\ 3\mathrm{p}_{3/2}^1$&$\color{blue}1\mathrm{s}^2\color{black}\ 2\mathrm{s}^1\ 3\mathrm{p}_{3/2}^1$  &  \SI{0.18}{}& Not Allowed &  \SI{1.09e-1}{}&{\SI{5.18e-1}{}}\\[2mm] 
\end{tabular}
\end{ruledtabular}
\end{table*}

It is worth mentioning that the maximum value obtainable for the resonance strength with and without GSA differs by more than three orders of magnitude in the interval $q=\interval{29+}{32+}$, due to the presence of the K-channels. The highest values of the $S_\mathrm{NEEC}$ in the L- and M-shells instead are comparable between the two cases. There are two main factors defining the final $S_\mathrm{NEEC}$ value for a given character of the nuclear transition and $E_\mathrm{n}$: (i) the resonance energy of the capture channel and (ii) the value of the microscopic NEEC rate $Y_\mathrm{NEEC}^{q,\alpha_0,\alpha_\mathrm{r}}$.
	(i) Because of the resonant nature of the NEEC process, $S_\mathrm{NEEC}$ increases dramatically when the energy released through electron capture nearly matches the nuclear transition.
	(ii) $Y_\mathrm{NEEC}^{q,\alpha_0,\alpha_r}$ depends on the overlap between the bound and free electron wave functions.
In the case of $^{73}$Ge, the enhancement found for the K-shell, compared to the highest value obtained under the GSA occurring for an L3 subshell, is due to an increase of the electron wavelength, since $Y_\mathrm{NEEC}^{q,\mathrm{K}}\leq Y_\mathrm{NEEC}^{q,\mathrm{L3}}$.

It is thus important to comment on the accuracy of the calculated energy levels.
In the Supplemental Material \cite{SM}, we compare the 38 energy levels available for Ge, obtained from the NIST website \cite{NIST_ASD}, and the same reproduced by FAC. The results show a good agreement with discrepancies between these levels usually smaller than \SI{1}{\electronvolt} and in all cases comparable with the accuracy reported for the E2 nuclear transition of $^{73}$Ge.
Although the $S_\mathrm{NEEC}$ values of the nearly-resonant energy levels are affected by the accuracy of FAC, 27 K-channels are present in the range $E_\mathrm{r}= [\SIrange[range-phrase ={, } ,range-units = single]{0}{10}{]\ \electronvolt}$ and 18 still forbidden in the range $E_\mathrm{r}= [\SIrange[range-phrase ={, } ,range-units = single]{-10}{0}{]\ \electronvolt}$. Thus, a shift of few eV does not affect our conclusions.
Similar screening effect on K-channels can be found in other isotopes as $^{98}$Tc and $^{125}$Te.
In the latter case, contrary to what happens for $^{73}$Ge, a further increase of $S_\mathrm{NEEC}$ is expected due to a higher value of $\alpha_\mathrm{IC}^\mathrm{q=0,K}$ compared to $\alpha_\mathrm{IC}^{\mathrm{q=0,L}i}$, with $i\in \{1,2,3\}$.

An increase of the resonance strength is particularly valuable when NEEC is compared to competitive processes, such as the direct-photoexcitation (DP) in the laser-generated plasma scenario \cite{gunst2015mutual,doolen1978nuclearC,gosselin2004enhanced,Gunst2018}.
Here, the discrimination of the two processes relies on the total number of excited nuclei, proportional to the corresponding photon or electron flux in plasma and the corresponding resonance strengths.
In a tabletop laser based setup, the photon flux can exceed the electron flux by several orders of magnitude \cite{SM, Gunst2018,Gibbon1996,Chung2005}.
This might hinder the observation of NEEC even for such promising nuclei as $^{73}$Ge, for which the DP resonance strength of the $E_2$ transition
is $S_\gamma=\SI{1.93e-6}{\barn\electronvolt}$, significantly smaller than the highest $S_{\mathrm{NEEC}}=\SI{1.25e-4}{\barn\electronvolt}$, obtained under the GSA.
Conversely, lifting the GSA allows for the appearance of
capture channels in the K-shell characterized by higher $S_\mathrm{NEEC}$ values.
This is particularly relevant if an additional external electron source is considered. For a few keV temperature plasma, the flux of electrons at low energies corresponding to K-channels is small and of the order of \SI{e21}{\per\square\centi\meter\per\second\per\electronvolt}. Under this condition, the use of an external adjustable electron source \cite{SM,FilippettoFLUX} could allow to overcome the deficit in the electron flux and decouple it from other plasma parameters. We point out here that the determination of the total number of excited nuclei would require the knowledge of the survival time of each atomic configuration in a specific experimental scenario. To the best of our knowledge, this level of detail is not available with current simulation tools.

In out-of-equilibrium scenarios, excited electronic configurations might be more likely to occur \cite{son2014quantum} and the same can hold true for the $^{93}$Mo isomer depletion of Ref. \onlinecite{Chiara2018}.
Indeed, during the entire impact with the carbon target, $^{93}$Mo ions are considered to be in their electronic ground state. This makes the contributions from the L-shell negligible, although the resonance strengths for the L-channels are the highest \cite{Wu2019}. This happens because the ion fraction in the charge state $q\geq 33+$ required for L-shell vacancies is extremely small when the resonant conditions are met.
Recently, a study considering the Compton profile of target electrons \cite{rzadkiewicz2021novel} shed new light on the importance of these L-channels, shifting upward by several orders of magnitude their theoretical contribution to the partial NEEC probability.
In particular, this study shows that the L-channels are no longer insignificant and their contribution is comparable with that of higher shells. Nevertheless, the total NEEC probability, accounting for charge state distribution and available vacancies, only slightly increases leaving the current discrepancy mostly unaltered. Indeed, under the GSA, the L-channels are not available at low projectile energies, where they give most of their contribution \cite{rzadkiewicz2021novel}. If instead electronic excitations would make L-vacancies survive even for $q<33+$, NEEC-EXI might reveal new capture L-channels even at low ion beam energy. The presence of these new channels, combined with their persistence over an energy continuum \cite{rzadkiewicz2021novel}, might possibly reduce the discrepancy between the experimental observation and theoretical predictions.

In NEEC scenarios only the energy matching between free electrons, bound states and nuclear transitions has been historically addressed.
Since selection rules for NEEC require $j_\mathrm{c}-L \leq j_\mathrm{f} \leq j_\mathrm{c}+L$, where $L$ is the multipolarity of the transition, in Ref. \onlinecite{Madan2020} we proposed that angular momentum matching could have given the possibility to select and enhance the capture in the innermost shells, by tuning the individual orbital angular momentum (OAM) $l\hbar$ of an external free electron beam \cite{verbeeck2010production,lloyd2017electron,verbeeck2018demonstration,vanacore2020spatio}, using phase plates or chiral plasmons \cite{bliokh2017theory,Vanacore2019,grillo2017measuring}. When using such OAM-carrying electrons (called vortex beam) the expressions used for $Y_\mathrm{NEEC}$ do not hold in the same form, thus leading to different values for the $S_\mathrm{NEEC}$. Recently, this has been shown in detail to be a way to increase the NEEC cross-section by several orders of magnitude \cite{wu2022dynamical}.
The combination of this additional degree of freedom with the presence of excited electronic configurations
could open a possibility to further boost the NEEC rate in a plasma scenario by providing specific atomic vacancies and pulsed vortex electrons at the resonant energy to selectively choose the capture in the desired shell.

In conclusion, we have shown that the common assumption that NEEC takes place in an ion in its electronic ground state significantly restricts the available channels. By lifting this condition, we have shown that in $^{73}$Ge the NEEC resonance strengths gain more than three orders of magnitude.
Thus, this work heralds the possibility of a reevaluation of the isotopes prematurely disregarded
and those already in use in out-of equilibrium scenarios.
These findings could open a new route for an externally controlled nuclear excitation by providing excited configurations and resonant engineered electrons from an external source, thus selecting the promising channels for on-demand isomer depletion. In particular, the inclusion of excited electronic configurations in the theoretical model describing the first NEEC observation in $^{93}$Mo, as here done for $^{73}$Ge,  could reduce the discrepancy between the actual theoretical predictions and experimental observation.

\begin{acknowledgments}
The LUMES laboratory acknowledges support from the NCCR MUST and Google Inc. The authors would like to thank G. M. Vanacore, C. J. Chiara, and J. J. Carroll for insightful discussions. The authors are indebted to A. Pálffy and Y. Wu for their useful feedback and for providing details on the derivation of the theory presented in Ref. \onlinecite{Gagyi-Palffy06}.
\end{acknowledgments}



\input{Manuscript.bbl}

\end{document}


\preprint{APS/123-QED}

\title{Supplemental Material: Nuclear Excitation by Electron Capture in Excited Ions}

\author{Simone Gargiulo}%
\email{simone.gargiulo@epfl.ch}

\author{Ivan Madan}
\author{Fabrizio Carbone}

\email{fabrizio.carbone@epfl.ch}
\affiliation{%
 Institute of Physics (IPhys), Laboratory for Ultrafast Microscopy and Electron Scattering (LUMES),\\
École Polytechnique Fédérale de Lausanne (EPFL), Lausanne 1015 CH, Switzerland
}%

\maketitle

\tableofcontents

\section{NEEC resonance strength derivation}
Formally, the NEEC cross-section can be written similarly to the cross-section for the formation of an excited compound nucleus, in which the entrance channel $\alpha$ is represented by the incoming electron. Thus, it is possible to write the NEEC cross section as the \textit{Breit-Wigner} one-level formula \cite{wong2008introductory, Zadernovsky2002, Harston}:

\begin{eqnarray}
\sigma_\mathrm{NEEC} = && \dfrac{\pi}{k_\mathrm{e}^2}\ \Gamma_\alpha\ \dfrac{\Gamma_\mathrm{r}}{(E-E_\mathrm{r})^2+\nicefrac{\Gamma_\mathrm{r}^2}{4}} \nonumber \\ = && \dfrac{2\pi^2}{k_\mathrm{e}^2}\ \Gamma_\alpha L_\mathrm{r}(E-E_\mathrm{r}),
\label{eq:NEECcs}
\end{eqnarray}
\noindent
where $k_\mathrm{e}$ is the electron wave-number, $\Gamma_\mathrm{r}$ is the natural resonance width given by the sum of the atomic ($\Gamma^\mathrm{At}$) and nuclear ($\Gamma_\mathrm{N}$) widths, $\Gamma_\alpha$ represents the transition width of the entrance channel $\alpha$ and $L_\mathrm{r}$ is a Lorentzian function centered at the resonance energy of the free electron $E_\mathrm{r}$.
 $\Gamma_\alpha$ is defined  by the microscopic NEEC reaction rate $Y_\mathrm{NEEC}$, via $\Gamma_\alpha= \hbar\ Y_\mathrm{NEEC} $:
\begin{equation}
 \sigma_{\mathrm{NEEC}}  = \dfrac{\lambda_\mathrm{e}^2}{2}\ \hbar\ Y_{\mathrm{NEEC}}^{q,\alpha_\mathrm{r}}(E)\, L_\mathrm{r}(E-E_\mathrm{r}) ,
\label{eq:NEEC-GSA}
\end{equation}
\noindent
 The integrated NEEC cross section, called resonance strength $S_{\mathrm{NEEC}}$, is then defined as:
\begin{equation}
 S_\mathrm{NEEC}^{q,\alpha_\mathrm{r}} = \int\  \mathrm{d}E\ \sigma_\mathrm{NEEC}(E)\ = \int \mathrm{d}E\ \dfrac{\lambda_\mathrm{e}^2}{2}\ \hbar\ Y_{\mathrm{NEEC}}\, L_\mathrm{r}(E-E_\mathrm{r})\, .
\label{eq:ResStrength}
\end{equation}
\noindent
\subsection{Internal Conversion Coefficients (ICC) scaling procedure}

The microscopic NEEC rate $Y_\mathrm{NEEC}$ \cite{Gunst2018} is related to the internal conversion rate ($A_\mathrm{IC}$) through the principle of detailed balance \cite{KineticPhotons}. Thus, in case of unpolarized beams we have:

\begin{equation}
Y_{\mathrm{NEEC}} = \dfrac{(2J_\mathrm{E}+1)(2j_\mathrm{c}+1)}{(2J_\mathrm{G}+1)(2j_\mathrm{f}+1)} A_\mathrm{IC}\,
\label{eq:detBalance}
\end{equation}
\noindent
where $J_\mathrm{E}$, $J_\mathrm{G}$ and $j_\mathrm{c}$ and $j_\mathrm{f}$ represent the nuclear spins of the excited and ground states and the total angular momenta of the captured and free electrons, respectively  \cite{Harston}. The NEEC cross section is then given by:
\begin{equation}
\sigma_{\mathrm{NEEC}} =\dfrac{(2J_\mathrm{E}+1)(2j_\mathrm{c}+1)}{(2J_\mathrm{G}+1)(2j_\mathrm{f}+1)}\ \dfrac{\lambda_\mathrm{e}^2}{2}\ \Gamma_{\mathrm{IC}}\ L_\mathrm{r}(E-E_\mathrm{r})\, ,
\label{eq:NEECcs3}
\end{equation}
\noindent
where $\lambda_e$ is the electron wavelength and $\Gamma_\mathrm{IC} = \hbar\ A_\mathrm{IC}$.
 More precisely, the microscopic NEEC and IC rate depend on the particular subshell $nl_j$ in which the electron is captured and charge state $q$ of the ion prior the electron capture. This information is condensed in the evaluation of the partial internal conversion coefficient $\alpha_\mathrm{IC}^{q,\alpha_\mathrm{r}}$, that depends on the final electronic configuration ($\alpha_\mathrm{r}$) and on the ion charge state $q$ prior to the electron capture.
Under the GSA, the initial electronic configuration ($\alpha_0$) is uniquely defined by the charge state $q$ --- i.e. the electronic ground state --- while the final electronic configuration ($\alpha_\mathrm{r})$ depends also on the particular capture channel. Thus, Eq. (\ref{eq:NEECcs3}) can be  expressed as:

\begin{equation}
\sigma_{\mathrm{NEEC}}^{q,\alpha_\mathrm{r}} =S\ \dfrac{\lambda_\mathrm{e}^2}{2}\ \alpha_\mathrm{IC}^{q,\alpha_\mathrm{r}}\ \Gamma_{\gamma}\ L_\mathrm{r}(E-E_\mathrm{r})\, ,
\label{eq:NEECcs4}
\end{equation}
\noindent
where $S=\nicefrac{(2J_\mathrm{E}+1)(2j_\mathrm{c}+1)}{(2J_\mathrm{G}+1)(2j_\mathrm{f}+1)}$ and $\Gamma_\gamma$ is the width of the electromagnetic nuclear transition.
ICCs  for  neutral  atoms  are  estimated  by  using  the frozen  orbital  (FO)  approximation based  on  the  Dirac-Fock calculations \cite{kibedi2008evaluation}, while for ionized atoms a linear scaling dependence is assumed \cite{Rzadkiewicz2019, rysavy2000reliability}:
\begin{equation}
\dfrac{\alpha_{\mathrm{IC}}^{q,\alpha_\mathrm{r}}}{E_\mathrm{b}^{q,\alpha_\mathrm{r}}}=\dfrac{\alpha_{\mathrm{IC}}^{q=0,nl_j}}{E_\mathrm{b}^{q=0,nl_j}} \Biggl( \dfrac{n_\mathrm{h}}{n_\mathrm{max}} \Biggr) \, ,
\label{eq:NEECcs5}
\end{equation}
\noindent
where $E_\mathrm{b}^{q,\alpha_0,\alpha_\mathrm{r}}$ and $E_\mathrm{b}^{{q=0,}nl_j}$ are the binding energies for ions in the charge state $q$ and neutral atoms, respectively. Their ratio accounts for the increase of the ICCs with the ionization level \cite{gosselin2004enhanced,rysavy2000reliability}.  The ratio between the present $n_\mathrm{h}$ and the maximum $n_\mathrm{max}$ number of holes in the capture subshell $nl_j$ accounts for the decrease of the ICCs for partially filled subshells \cite{rysavy2000reliability}. The binding energies for neutral atoms were taken from tables \cite{larkins1977semiempirical}, while the ones for highly ionized atoms are calculated  with FAC  \cite{gu2008flexible}, obtained as energy difference between the initial ($\alpha_\mathrm{0}$) and final electronic configurations ($\alpha_\mathrm{r}$). The resonance energy $E_\mathrm{r}$ is then obtained as $E_\mathrm{n}-E_\mathrm{b}$ for the specific channel considered.
 Accuracy of these levels is assessed to be in the order of few eV \cite{massacrier2012extensive}.
As described in the main text, when NEEC is occurring in excited ions, the coefficient $\alpha_{\mathrm{IC}}^{q,\alpha_\mathrm{r}}$ also depends on the particular initial electronic configuration $\alpha_0$ and it has to be expressed as $\alpha_{\mathrm{IC}}^{q,\alpha_0,\alpha_\mathrm{r}}$:

\begin{equation}
\dfrac{\alpha_{\mathrm{IC}}^{q,\alpha_0,\alpha_\mathrm{r}}}{E_\mathrm{b}^{q,\alpha_0,\alpha_\mathrm{r}}}=\dfrac{\alpha_{\mathrm{IC}}^{{q=0,}nl_j}}{E_\mathrm{b}^{{q=0,}nl_j}}\ \Biggl(\dfrac{n_\mathrm{h}}{n_\mathrm{max}}\Biggr)\ ,
\label{eq:linear_scaling}
\end{equation}
\noindent
As a consequence, also the NEEC cross-section and the resonance strength have to be represented as $\sigma_\mathrm{NEEC}^{q,\alpha_0,\alpha_\mathrm{r}}$ and  $S_\mathrm{NEEC}^{q,\alpha_0,\alpha_\mathrm{r}}$, respectively.
When approaching the threshold $E_\mathrm{r} \leq  \SI{1}{\kilo\electronvolt}$, to account for the non-linear trend of the ICCs, Eq. \ref{eq:linear_scaling} is replaced with a non-linear model. For K-channels a fourth-order polynomial has been used, as similarly done in Ref. \onlinecite{kantele1989simplified}, to extrapolate the value of $\alpha_{IC}^{q=0,nl_j}$ as function of the energy. This non-linear fit replaces Eq. \ref{eq:linear_scaling}, as follow:

\begin{equation}
\alpha_{\mathrm{IC}}^{q,\alpha_0,\alpha_\mathrm{r}}=\alpha_{\mathrm{IC}}^{{q=0,}nl_j} \Biggl(E_\mathrm{n}-\bigr(E_\mathrm{b}^{q,\alpha_0, \alpha_\mathrm{r}}-E_\mathrm{b}^{q=0,K}\bigr)\Biggr)\ \Biggl(\dfrac{n_\mathrm{h}}{n_\mathrm{max}}\Biggr)\ ,
\label{eq:non-linear_scaling}
\end{equation}
\noindent
where the energies inside the parenthesis are the argument of the ICC. To improve comparison with the wavefunction formalism presented in Ref. \onlinecite{Palffy2006} we used $S=\nicefrac{(2J_\mathrm{E}+1)(2j_\mathrm{c}+1)}{(2J_\mathrm{G}+1)}$ as statistical factor given that different values of $j_\mathrm{f}$ contribute to the final $\alpha_\mathrm{IC}^{q,\alpha_\mathrm{r}}$.

\subsection{Wavefunction Formalism}
The relativistic bound and continuum wavefunctions, solution of the Dirac equations, are written as:

\begin{equation}
    \Psi_{n_b \kappa_b m_b}(\bf{r})= \dfrac{1}{r} \Bigg( \begin{matrix} P_{n_b\kappa_b} (r) \chi_{\kappa}^m(\theta,\phi)\\ iQ_{n_b\kappa_b} (r) \chi_{-\kappa}^m(\theta,\phi)\\
    \end{matrix}\Bigg)\
\end{equation}

\begin{equation}
    \Psi_{E\kappa m}(\bf{r})= \dfrac{1}{r} \Bigg( \begin{matrix} P_{E\kappa} (r) \chi_{\kappa}^m(\theta,\phi)\\ iQ_{E\kappa} (r) \chi_{-\kappa}^m(\theta,\phi)\\
    \end{matrix}\Bigg)\
\end{equation}
\noindent
where, $P(r)=r g(r)$ and $Q(r)=r f(r)$ are the Dirac spin orbitals, while $f(r)$ and g(r) are the large and small radial components.
As shown in Ref. \onlinecite{Palffy2006}, in the case of an electric transition the NEEC rate can be written in atomic unit as:

\begin{equation}
    Y_n^{(e)L}= \dfrac{1}{4\pi\alpha} \dfrac{4\pi^2 \rho_i}{(2L+1)^2 } {B\! \uparrow} (2 j_b +1) \sum_\kappa |\widetilde{R}_{L,\kappa_b,\kappa}|^2 C(j_b\ L\ j; 1/2\ 0\ 1/2)^2
    \label{eq:NEEC_RATE}
\end{equation}
\noindent
In Eq. \ref{eq:NEEC_RATE} ${B\! \uparrow}$ is the reduced transition probability of the L\textsuperscript{th} multipolar transition, $C (j_1\ j_2\ j;\ m_1\ m_2\ m)$ is the Clebsch-Gordan coefficient. $j_\mathrm{b}$ and $j$ are  the total angular momentum, while $k_\mathrm{b}$ and $k$ are the Dirac angular momentum of the bound and free electron, respectively. $\kappa$ assumes only the values allowed by selection and parity rules.
In this formalism $j_\mathrm{b}$ and $j$ are equivalent to $j_\mathrm{c}$ and $j_\mathrm{f}$, respectively.
$\widetilde{R}_{L,k_b,k}$ is the radial integral and its expression is
\begin{equation}
    \widetilde{R}_{L,\kappa_b,\kappa} = \int_0^{+\infty} dr\ r^{-L-1} [P_{\kappa_b}(r)P_{E \kappa}(r) + Q_{\kappa_b}(r)Q_{E \kappa}(r) ]
    \label{eq:radial integral}
\end{equation}
and depends on the bound ($P_{\kappa_b}$,$Q_{\kappa_b}$) and free ($P_{E \kappa}$,$Q_{E \kappa}$) electron wavefunctions, which are obtained as solutions of the Dirac equations for a particular atomic configuration using FAC \cite{gu2008flexible}. Compared to Ref. \onlinecite{Palffy2006} a different normalization of the free wavefunctions is performed by FAC, thus $P_{E\kappa}$ and $Q_{E\kappa}$ have to be furtherly normalized to $\sqrt{\pi}$.

\subsection{Atomic Widths for K-channels}
In case of K-channels the width of the resonant process $\Gamma_\mathrm{NEEC}=\Gamma_{nl}^\mathrm{At}+\Gamma_\mathrm{N}\approx \Gamma_{nl}^\mathrm{At}$ can be comparable with the resonance energy $E_\mathrm{r}$. Thus, $Y_\mathrm{NEEC}$ can not be moved out of the integral in Eq. \ref{eq:ResStrength}. Atomic widths for all the specific 100 excited electronic configurations leading to a K-capture have been evaluated using the XATOM code \cite{jurek2016xmdyn, son2011impact, murphy2014femtosecond}. Integral have been performed over an interval of $8\times  \Gamma_{nl}^\mathrm{At}$ in all the cases, apart from the first two channels where the lower limit for the integration has been set to \SI{0.01}{\electronvolt}. This choice does not affect the main result of NEEC-EXI since the third and fourth channels at \SI{0.98}{\electronvolt} and \SI{1.15}{\electronvolt}, not affected by the truncation of the lower limit of the integral, have resonance strengths comparable to the first two.

\section{Recoupling coefficients}
In NEEC-EXI, for a given charge state $q$, electrons are assigned to a particular shell from the innermost to the outermost (K, L, M) encompassing all possible combinations (e.g., for 3 electrons all the cases between $1s^2\ 2s^1$ and $3d_{5/2}^3$ are considered; for 4 electrons all the cases between $1s^2\ 2s^2$ and $3d_{5/2}^4$).  All these electronic states are used as initial configurations $\alpha_0$.

When considering the electron capture in this context, it has to be taken into account the possible recoupling between the electron involved in the capture and those that are already in the atom.
The case of initially fully ionized atom is trivial since no recoupling is occurring.
In case the capture leads to the formation of an ion having two electrons, the selection rules are satisfied if the spectator electron (not involved in NEEC)
preserves its orbital angular momentum during the process.
The NEEC cross-section would be non-zero only if this condition is satisfied.
For the case of three electrons, we can recognize two situations: (i) the capture does not break the coupling or (ii) the capture breaks the coupling. With $j_1$, $j_2$ and $j_\mathrm{c}$ we denote the total angular momenta of the two spectator electrons and the one involved in the capture, respectively. Case (i) occurs when \cite{Bilous2017}:
\begin{enumerate}[noitemsep,topsep=0pt]
    \item $j_1$ and $j_2$ firstly couple to $J_{12}$,
    \item $J_{12}$ then couples with $j_\mathrm{c}$ forming $J$,
\end{enumerate}
\noindent
where $J_{12}$ is the initial angular momentum, not broken by the capture.
The NEEC cross-section can be considered non-zero only when the orbital angular momenta of the two spectator electrons and their coupling (thus $j_1,j_2,J_{12}$) remain unchanged among the initial and final states.
The other possibility is (ii):
\begin{enumerate}[noitemsep,topsep=0pt]
  \setlength\itemsep{0em}
    \item $j_2$ and $j_\mathrm{c}$ firstly couple to $J_\mathrm{2c}$,
    \item $J_\mathrm{2c}$ then couples with $j_1$ forming $J$.
\end{enumerate}
\noindent
Here, the capture breaks the initial coupling and the expression of the resonance strength in Eq. \ref{eq:ResStrength} has an additional coefficient \cite{yutsis1962mathematical,biedenharn1981racah,Bilous2017}:
\begin{equation}
\Lambda =\abs{ \braket{[j_1,(j_2,j_\mathrm{c})J_\mathrm{2c}];J|[(j_1,j_2)J_{12},j_\mathrm{c}];J}}^2
=(2\cdot J_{12}+1)(2\cdot J_\mathrm{2c}+1) \sixj{j_1}{j_2}{J_{12}}{j_\mathrm{c}}{J}{J_\mathrm{2c}}^2\ ,\label{eq:Coeff_3electrons}
\end{equation}
\noindent
where $\Lambda$ expresses the probability that a system with a coupling scheme defined by the bra vector $\bra{[j_1,(j_2,j_\mathrm{c})J_\mathrm{2c}];J}$ will be found in the scheme $\ket{[(j_1,j_2)J_{12},j_\mathrm{c}];J}$ \cite{biedenharn1981racah}. Whether a peculiar coupling is possible or not depends on the Wigner 6j-symbol. Notice that here $J_{12}$ also represents the total orbital angular momentum of the two-electron atomic system \textit{before} the electron capture.
With three electrons, three nontrivial coupling schemes exist and as soon as more electrons ($n$) are added to the ion, the number of possible couplings increases as $(2n-3)!!$  \cite{biedenharn1981racah}.

Indeed, in case of four electrons it is possible to distinguish fifteen nontrivial coupling schemes. Considering an initial electronic configuration $\alpha_0$ having a charge state $q=(Z-3)+$, with all the three electrons belonging to different orbitals, the electron capture (EC) of a fourth electron can lead to various final configurations. In this circumstances, it is possible to distinguish the following substantially different scenarios:

\begin{itemize}
    \item $2\mathrm{p}_{1/2}^1\ 3\mathrm{s}^1\ 3\mathrm{d}_{5/2}^1$ $\rightarrow$ $\color{blue} 1\mathrm{s}^1 \color{black} \
    2\mathrm{p}_{1/2}^1 \ 3\mathrm{s}^1\ 3\mathrm{d}_{5/2}^1$\, ;
    \item $1\mathrm{s}^1\ 3\mathrm{s}^1\ 3\mathrm{d}_{5/2}^1$ $\rightarrow$ $1\mathrm{s}^1\ \color{blue} 2\mathrm{p}_{1/2}^1  \color{black} \ 3\mathrm{s}^1\ 3\mathrm{d}_{5/2}^1$\, ;
    \item $1\mathrm{s}^1\ 2\mathrm{p}_{1/2}^1\ 3\mathrm{d}_{5/2}^1$ $\rightarrow$ $1\mathrm{s}^1\ 2\mathrm{p}_{1/2}^1\ \color{blue}  3\mathrm{s}^1 \color{black} \ 3\mathrm{d}_{5/2}^1$\ \, ;
    \item $1\mathrm{s}^1\ 2\mathrm{p}_{1/2}^1\ 3\mathrm{s}^1$ $\rightarrow$ $1\mathrm{s}^1\ 2\mathrm{p}_{1/2}^1\ 3\mathrm{s}^1\ \color{blue} 3\mathrm{d}_{5/2}^1\color{black}$ \, .
\end{itemize}
\noindent
The recoupling coefficients associated with the first three cases are:

\begin{widetext}
\begin{gather}
 \braket{[(j_1,j_2)J_{12},j_3]J_{123},j_4;J|[(j_2,j_3)J_{23},j_4]J_{234},j_1;J}=(-1)^{\theta_1} R_1 \sixj{j_1}{j_2}{J_{12}}{j_3}{J_{123}}{J_{23}} \sixj{j_1}{J_{23}}{J_{123}}{j_4}{J}{J_{234}}\ \label{eq:Case5.1},
\\
R_1=\sqrt{(2\cdot J_{12}+1)(2\cdot J_{123}+1)(2\cdot J_{23}+1)(2\cdot J_{234}+1)}\ ,\nonumber
\end{gather}
\begin{gather}
 \braket{[(j_1,j_2)J_{12},j_3]J_{123},j_4;J|[(j_1,j_3)J_{13},j_4]J_{134},j_2;J}=(-1)^{\theta_2} R_2 \sixj{j_2}{j_1}{J_{12}}{j_3}{J_{123}}{J_{13}} \sixj{j_2}{J_{13}}{J_{123}}{j_4}{J}{J_{134}}\ \label{eq:Case5.2},
\\
R_2=\sqrt{(2\cdot J_{12}+1)(2\cdot J_{123}+1)(2\cdot J_{13}+1)(2\cdot J_{134}+1)}\ ,\nonumber
\end{gather}
\begin{gather}
 \braket{[(j_1,j_2)J_{12},j_3]J_{123},j_4;J|[(j_1,j_2)J_{12},j_4]J_{124},j_3;J}=(-1)^{\theta_3} R_3 \sixj{j_3}{J_{12}}{J_{123}}{j_4}{J}{J_{124}}\ \label{eq:Case5.3} ,
\\
R_3=\sqrt{(2\cdot J_{123}+1)(2\cdot J_{124}+1)}\ , \nonumber
\end{gather}
\end{widetext}
\noindent
respectively. In these equations, the phase factor --- indicated by $\theta_i$ --- is not reported since it is irrelevant in the evaluation of the probability $\Lambda=\abs{\braket{\textbf{a}|\textbf{b}}}^2$, with \textbf{a} and \textbf{b} being the two state vectors expressing the coupling.
Differently, the selection rule for the capture in the outermost shell requires that the electrons not involved in NEEC conserve their individual $j_i$ and their intermediate couplings $J_{ik}$, $J_{ikl}$.

\begin{figure}[!h]
\includegraphics[scale=0.75]{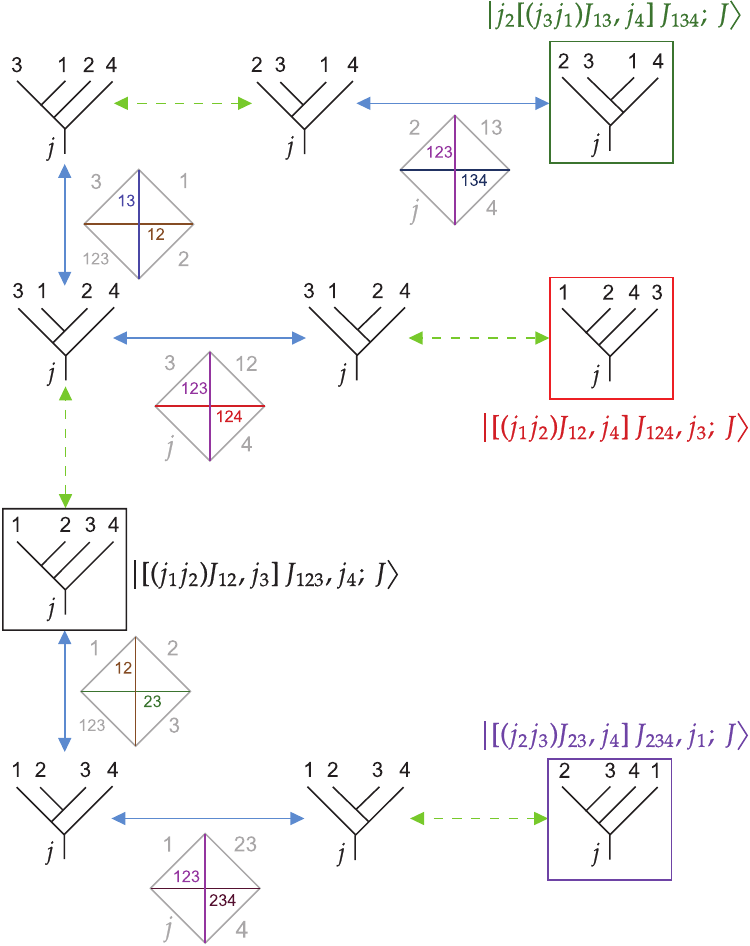}
\caption{\label{fig:binary_trees} Binary trees connecting the three coupling schemes given in Eqs. \ref{eq:Case5.1}-\ref{eq:Case5.3}. Starting from one state, applying a sequence of the elementary operations presented, it is possible to arrive at the desired final state.
\textit{Exchange} and \textit{flop} operations are represented by green dashed and blue solid lines, respectively. The rhombuses graphically represent, in Yutsis notation \cite{yutsis1962mathematical}, the Wigner 6-j symbols for the considered \textit{flop} operation. Eq. \ref{eq:Case5.1}, for example,  can be obtained following the path going from the state $\ket{[(j_2,j_3)j_4]j_1}$, boxed in purple, to the state vector $\ket{[(j_1,j_2)j_3]j_4}$, indicated by a black square, as multiplication of two Wigner 6-j symbols (with the relative square root terms $R_i$ due to the two \textit{flop} operations) and an additional phase factor coming from the \textit{exchange} operation.
Eq. \ref{eq:Case5.2} is given by the path going from the green to the black boxes, while Eq. \ref{eq:Case5.3} is the path connecting the red and black boxes.}
\end{figure}

These recoupling coefficients, reported in Eqs. \ref{eq:Case5.1}-\ref{eq:Case5.3}, can be understood by graphical means using Yutsis notation \cite{yutsis1962mathematical} and binary trees \cite{biedenharn1981racah,Aquilanti2008}, presented in Fig. \ref{fig:binary_trees}.  Each pair of binary trees, whose leaves are labelled with the four uncoupled angular momenta, can be connected by two types of elementary operations: \textit{exchange} and \textit{flop} \cite{Fack1995,Fack1999,biedenharn1981racah}.
An \textit{exchange}, represented by a dashed arrow, does not lead to a rearrangement of the orbital angular momenta, but to a swap of the $j_i$ around one node. Thus, the relative transformation coefficient corresponds to a phase factor. The \textit{flop} operation instead, shown as a solid arrow, is effectively a recoupling relating two trees with two alternative nets connecting the leaves.
This latter transformation is defined by a Racah coefficient, proportional to a \textit{6j-symbol} (represented by the rhomboidal Yutsis graph in Fig. \ref{fig:binary_trees}).

Electron capture can occur in many other circumstances. As for example, we can have an initial configuration $\alpha_0$ with $q=(Z-3)+$ but with two out of three electrons belonging to the same orbital:

\begin{itemize}
    \item  $2\mathrm{p}_{1/2}^1\ 3\mathrm{d}_{5/2}^{\color{red}2\color{black}}$ $\rightarrow$ $\color{blue}1\mathrm{s}^1 \color{black}\ 2\mathrm{p}_{1/2}^1\ 3\mathrm{d}_{5/2}^{\color{red}2\color{black}}$\, ;
    \item $2\mathrm{p}_{1/2}^1\ 3\mathrm{d}_{5/2}^{\color{red}2\color{black}}$ $\rightarrow$ $2\mathrm{p}_{1/2}^1\ \color{blue}2\mathrm{p}_{3/2}^1 \color{black}\ 3\mathrm{d}_{5/2}^{\color{red}2\color{black}}$\, ;
    \item $2\mathrm{p}_{1/2}^1\ 3\mathrm{d}_{5/2}^{\color{red}2\color{black}}$ $\rightarrow$ $2\mathrm{p}_{1/2}^1\ \color{blue}3\mathrm{d}_{5/2}^3$\, .
\end{itemize}
\noindent
In Eqs. \ref{eq:Case5.4.1}-\ref{eq:Case6.2.4} we report the associated recoupling coefficients.

\begin{widetext}
\begin{gather}
 \braket{[(j_1,j_2)J_{12},(j_3,j_4)J_{34}];J|[j_2,(j_3,j_4)J_{34}]J_{234},j_1;J}=(-1)^{\theta_4} R_4 \sixj{j_1}{j_2}{J_{12}}{J_{34}}{J}{J_{234}} \label{eq:Case5.4.1},
\\
R_4=\sqrt{(2\cdot J_{12}+1)(2\cdot J_{234}+1)}\ ,\nonumber
\end{gather}
\begin{gather}
 \braket{[(j_1,j_2)J_{12},(j_3,j_4)J_{34}];J|[j_1,(j_3,j_4)J_{34}]J_{134},j_2;J}=(-1)^{\theta_5} R_5 \sixj{j_2}{j_1}{J_{12}}{J_{34}}{J}{J_{134}} \label{eq:Case5.4.2},
\\
R_5=\sqrt{(2\cdot J_{12}+1)(2\cdot J_{134}+1)}\ ,\nonumber
\end{gather}
\begin{gather}
 \braket{j_1,[(j_2,j_3)J_{23},j_4]J_{234};J|[j_1,(j_2,j_3)J_{23}]J_{123},j_4;J}=(-1)^{\theta_6} R_6 \sixj{j_4}{J_{23}}{J_{234}}{j_1}{J}{J_{123}} \label{eq:Case6.2.4},
\\
R_6=\sqrt{(2\cdot J_{234}+1)(2\cdot J_{123}+1)}\ ,\nonumber
\end{gather}
\end{widetext}
\noindent

Another relevant case is given by the initial electronic configuration ($\alpha_0$) of the type $1\mathrm{s}^12l^13l^1$ with $q=(Z-3)+$.
In fact, a final configuration ($\alpha_\mathrm{r}$) resulting from a capture in the K-shell would lead to the following scenario:

\begin{itemize}
    \item
$1\mathrm{s}^1\ 2\mathrm{p}_{1/2}^1\ 3\mathrm{d}_{3/2}^1$ $\rightarrow$ $\color{blue}1\mathrm{s}^2\color{black}\  2\mathrm{p}_{1/2}^1\ 3\mathrm{d}_{3/2}^1$ .
\end{itemize}
This is an example of the 100 K-capture channels identified in the manuscript. The associated recoupling coefficient is:

\begin{widetext}
\begin{gather}
 \braket{[(j_1,j_2)J_{12},j_3]J_{123},j_4;J|[(j_1,j_3)J_{13},j_4]J_{134},j_2;J}=(-1)^{\theta_{7}} R_{7} \sixj{j_2}{j_{1}}{\cancel{J_{12}}\color{red}^0}{j_{3}}{J_{123}}{J_{13}}  \sixj{j_2}{J_{13}}{J_{123}}{j_{4}}{J}{J_{134}} \label{eq:Case7.1.3},
\\
R_{7}=\sqrt{(2\cdot J_{12}+1)(2\cdot J_{123}+1)(2\cdot J_{13}+1)(2\cdot J_{134}+1)}\ =\nonumber \sqrt{(2\cdot J_{123}+1)(2\cdot J_{13}+1)(2\cdot J_{134}+1)}\, .\nonumber
\end{gather}
\end{widetext}

\noindent

\clearpage
\section{NEEC resonance strength for $^{73}$Ge under Ground state assumption}

\begin{figure*}[b]
\includegraphics[scale=0.27]{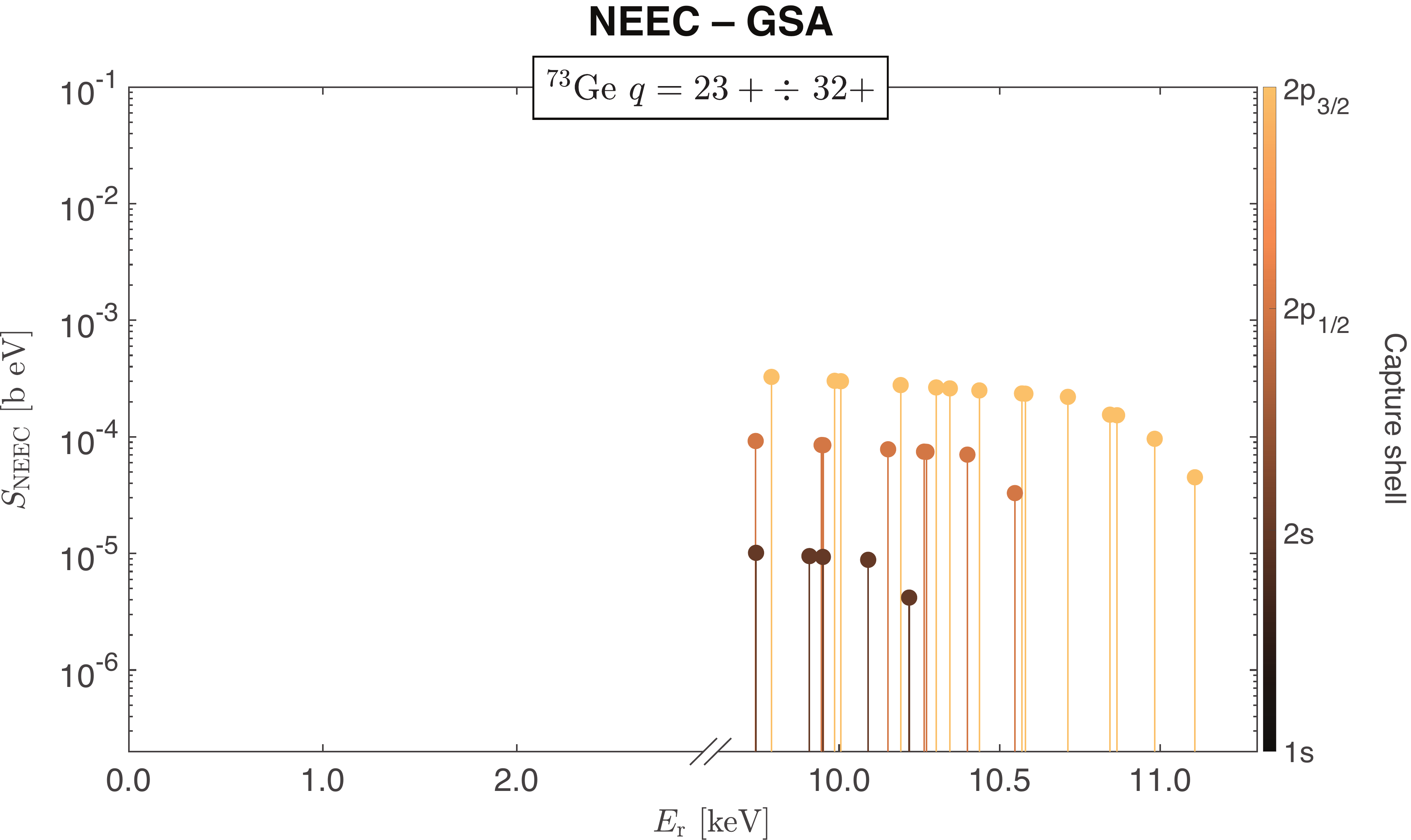}
\caption{\label{fig:Ge73} Resonance strengths for the L-shell channels of $^{73}$Ge, under the GSA. Charge state $q$ ranges between 32+ (bare nucleus) to $q=23+$ (after the electron capture the K and L shells are both closed).}
\end{figure*}

Considering the framework here presented, Eq. \ref{eq:ResStrength} leads to the resonance strengths shown  in Fig. \ref{fig:Ge73} and in Table \ref{tab:Ge73} for $^{73}$Ge.
Here, only the capture up to the L-shells has been considered, while the charge state varied between the one of the bare nucleus ($q=Z+$) all the way down to the closure of the K- and L-shells ($q=(Z-8)+$). The direct consequence of using GSA is the fact that for
$^{73}$Ge only 27 L-channels exist in the interval  $q= \interval{23+}{32+}$, since NEEC in the K-shell is energetically forbidden.

\begin{table}[!h]
\caption{\label{tab:Ge73}
NEEC resonance strength $S_\mathrm{NEEC}$, partial internal conversion coefficient $\alpha_\mathrm{IC}^{q,\alpha_\mathrm{r}}$ and the energy of the continuum electron $E_\mathrm{r}$ for various charge state $q$ (before the capture) and capture shells $\alpha_\mathrm{r}$, in case of $^{73}$Ge. $^{73}$Ge has its first excited state at \SI{13.2845}{\kilo\electronvolt}. Here, the symbol for the final electronic configuration $\alpha_\mathrm{r}$ is used to indicate the capture channel, as they are uniquely connected once the charge state $q$ is assigned.}
\begin{ruledtabular}
\begin{tabular}{ccccc}
 {$q$}&{$\alpha_\mathrm{r}$}&{$E_\mathrm{r}$ [keV]}&{$\alpha_\mathrm{IC}^{q,\alpha_\mathrm{r}}$}&{$S_\mathrm{NEEC}$ [b$\cdot$eV]}\\[0.1cm]
\hline\\[0.01mm]
{32+}&{$2\mathrm{s}_{1/2}$}&{$9.74$}&{$72.02$}&{$\SI{1.01e-5}{}$}\\
{32+}&{$2\mathrm{p}_{1/2}$}&{$9.74$}&{$655.55$}&{$\SI{9.21e-5}{}$}\\
{32+}&{$2\mathrm{p}_{3/2}$}&{$9.79$}&{$1167.65$}&{$\SI{3.26e-4}{}$}\\
{31+}&{$2\mathrm{p}_{1/2}$}&{$9.95$}&{$617.59$}&{$\SI{8.50e-5}{}$}\\
{31+}&{$2\mathrm{p}_{3/2}$}&{$9.99$}&{$1101.94$}&{$\SI{3.02e-4}{}$}\\
{30+}&{$2\mathrm{s}_{1/2}$}&{$10.09$}&{$64.93$}&{$\SI{8.82e-6}{}$}\\
{30+}&{$2\mathrm{p}_{1/2}$}&{$10.15$}&{$579.29$}&{$\SI{7.82e-5}{}$}\\
{29+}&{$2\mathrm{p}_{3/2}$}&{$10.30$}&{$996.35$}&{$\SI{2.65e-4}{}$}\\
{26+}&{$2\mathrm{p}_{3/2}$}&{$10.71$}&{$859.36$}&{$\SI{2.20e-4}{}$}\\
{25+}&{$2\mathrm{p}_{3/2}$}&{$10.84$}&{$611.68$}&{$\SI{1.55e-4}{}$}\\
{23+}&{$2\mathrm{p}_{3/2}$}&{$11.11$}&{$181.81$}&{$\SI{4.50e-5}{}$}
\end{tabular}
\end{ruledtabular}
\end{table}

\clearpage
\section{Accuracy of the energy levels computed with the Flexible Atomic Code (FAC)}
\begin{figure*}[!h]
\includegraphics[width=\linewidth]{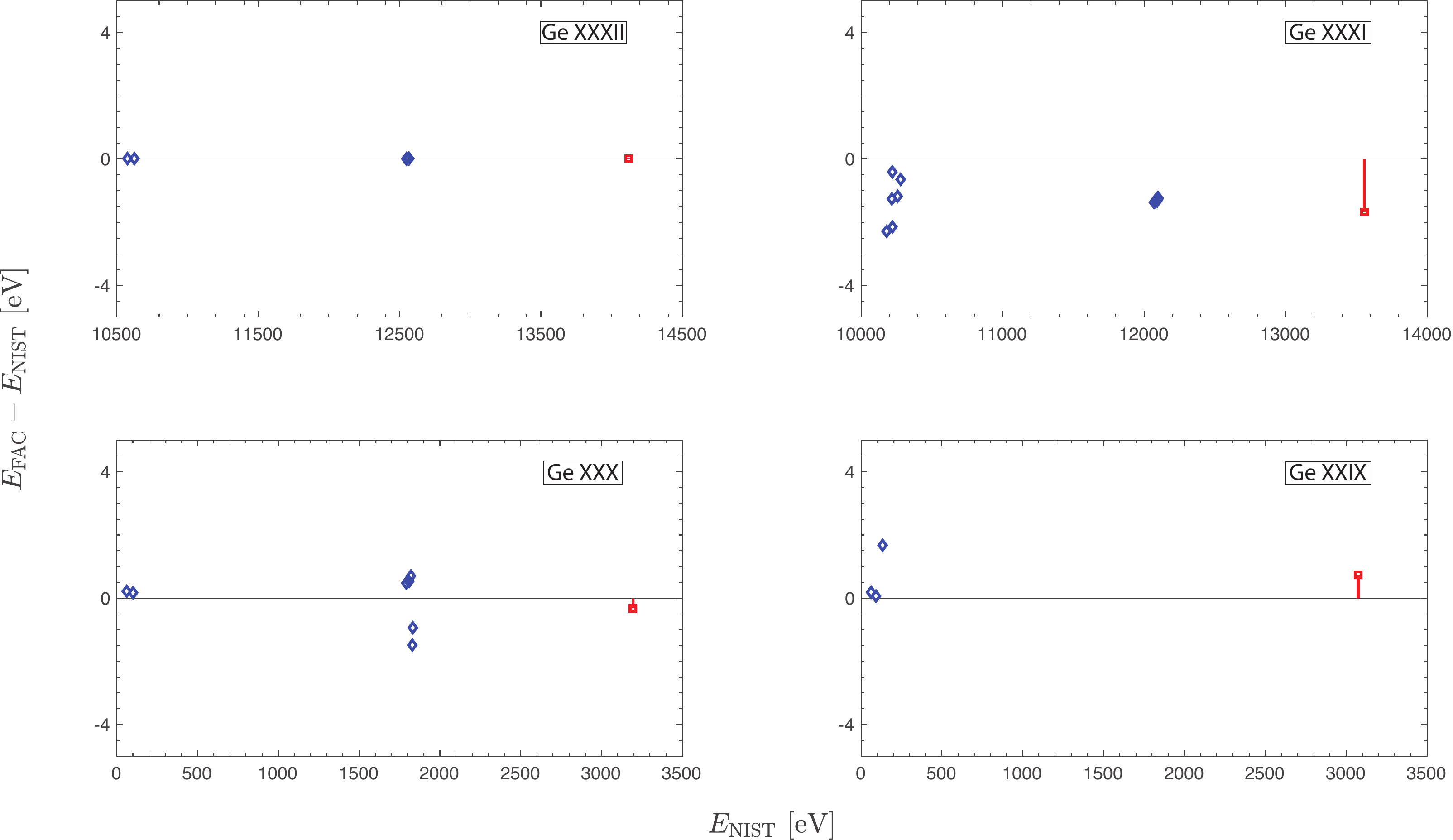}
\caption{\label{fig:Accuracy_FAC} Energy differences between the levels computed with FAC and the same energy levels available from the NIST database \cite{NIST_ASD} for the charge state interval $q=\interval{29+}{32+}$ of Ge. Differences between excited electronic states are shown as blue solid lines, while in red the differences between ionization energies.}
\end{figure*}

For the four ions of Ge, i.e. $q=\interval{29+}{32+}$, and the generic electronic configurations considered in the main text, our computation with the Flexible Atomic Code (FAC) \cite{gu2008flexible} leads to a total of 4565 energy levels. For the same ions and shells involved, the NIST database \cite{NIST_ASD} contains only 38 energy levels corresponding to 34 electronic excited states and 4 ionization energies. Thus, less than $1\%$ of the total energy levels reproduced by FAC can be compared with NIST data. This comparison is shown in Fig. \ref{fig:Accuracy_FAC} as energy differences between the levels computed with FAC and those obtained from the NIST database as a function of the NIST energy levels, as similarly done in Ref. \onlinecite{massacrier2012extensive}. The correspondence between the FAC and the NIST levels is established in terms of electronic configuration and total angular momentum.

\noindent
Agreement is excellent for the Hydrogen-like ion Ge XXXII, where the standard deviation of the energy differences is of $\SI{6.4}{\milli\electronvolt}$. The mean value of the energy differences is of $\SI{13.5}{\milli\electronvolt}$.  For this ion, the 9 levels reported in the NIST database are also theoretically evaluated. For the ion Ge XXXI is observed the maximum discrepancy. Here, the average differences for the 17 levels compared is of $\SI{1.32}{\electronvolt}$, while their standard deviation is of $\SI{0.44}{\electronvolt}$. The largest discrepancy of $\SI{-2.38}{\electronvolt}$ is for the $1\mathrm{s}^12\mathrm{s}^1$ ($^3\mathrm{S}_1$) level. 14 of these NIST levels are extrapolated or interpolated starting from the two known experimental values.
A good agreement, with a standard deviation of \SI{0.78}{\electronvolt} and an average value of the energy differences of $\SI{-0.08}{\electronvolt}$, is found also for the Ge XXX ion. Apart from the $1\mathrm{s}^2 3\mathrm{d}^1$ ($^2\mathrm{D}_{3/2}$) level, all the energy differences are $<$ \SI{1}{\electronvolt}. The $^2\mathrm{D}_{3/2}$ level reports a discrepancy of $\SI{-1.48}{\electronvolt}$. The 7 energy levels reported on the NIST database for this ion are experimentally observed, while the ionization energy is theoretically predicted.
For Ge XXIX, only 4 suitable levels are present on the NIST database. The mean value of the energy difference is of $\SI{0.67}{\electronvolt}$, while the standard deviation is of $\SI{0.73}{\electronvolt}$, mainly given by the discrepancy observed for the 1s$^2$2s$^1$2p$^1$ ($^1$P$_1$) level, that is of $\SI{1.68}{\electronvolt}$. Here, the 3 electronic excited states are experimentally observed, while the ionization energy is obtained from extrapolation.

The quality of the FAC calculations can be appreciated when discrepancies here reported are compared with the energy spread over which the studied electronic configurations persist, i.e. of few \SI{}{\kilo\electronvolt}.

\clearpage
\section{NEEC in Excited $^{73}$Ge ions}
\begin{figure*}[!h]
\includegraphics[width=\linewidth]{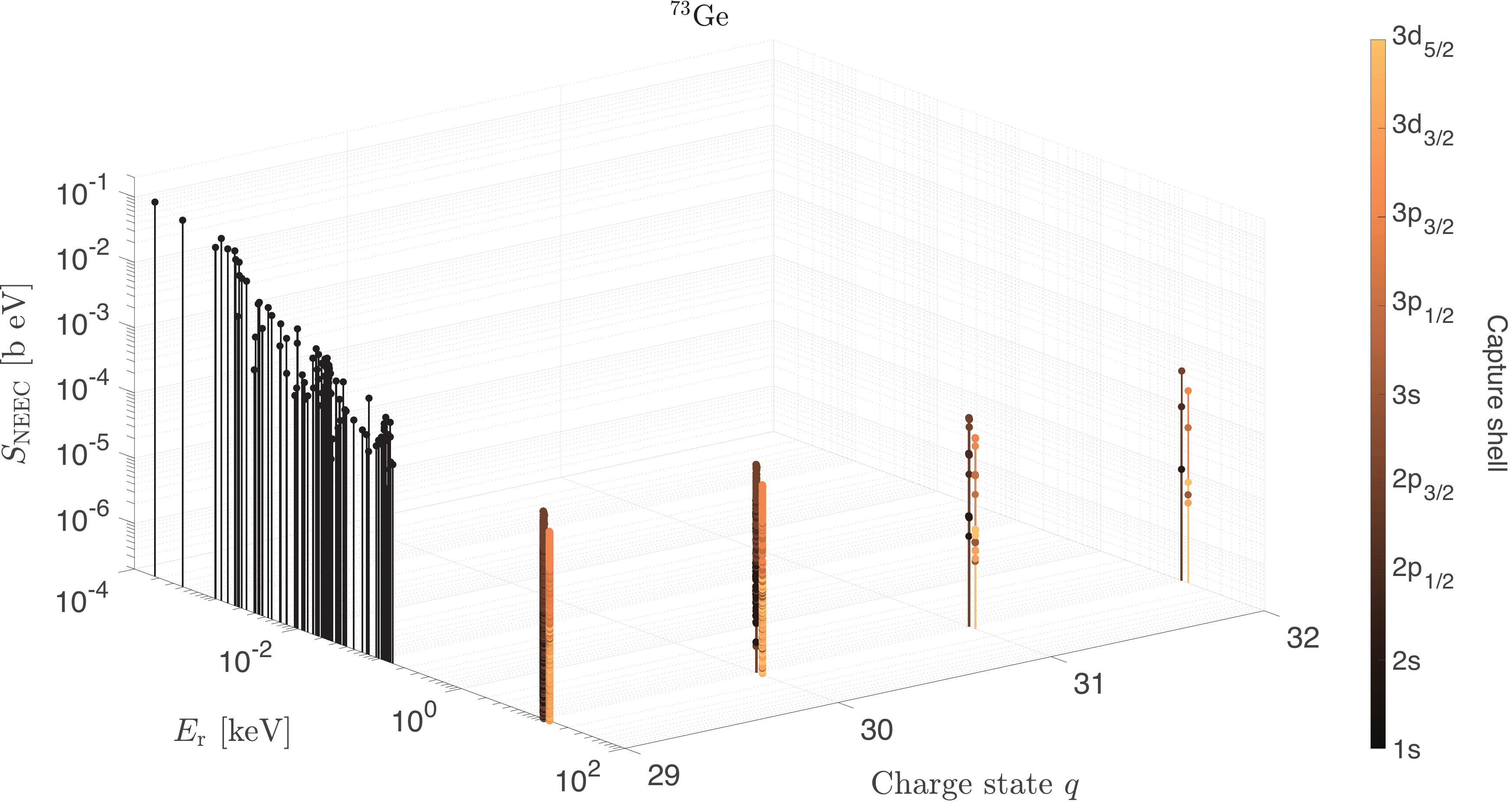}
\caption{\label{fig:Ge73_3D} Resonance strengths for $^{73}$Ge with $q= \interval{29+}{32+}$, in case all the possible combinations of initial ($\alpha_\mathrm{0}$) and final ($\alpha_\mathrm{r}$) electronic configurations are taken into account, as a function of the charge state $q$ and of the resonance energy $E_\mathrm{r}$. Each resonant channel is represented by a solid line, with its colors indicating the capture orbital, ranging from the K to the M shell. It is clear that when $q= 29+$, $E_\mathrm{b}^\mathrm{K}$ becomes smaller than $E_\mathrm{n}$ and nuclear excitation induced by a K-capture is possible. $q$ indicates the ion charge state before the electron capture.}
\end{figure*}

As described in the main text, lifting the GSA provides a total of 32823 capture channels in the charge state interval $q=\interval{29+}{32+}$, considering K-, L- and M-shells.
From Fig. \ref{fig:Ge73_3D} it is possible to observe that, upon filling the orbitals, electron screening lowers the energy release through the electron capture in the K-shell ($E_\mathrm{b}^\mathrm{K}$), till it becomes smaller than $E_\mathrm{n}$, and NEEC into the K shell starts to be possible. In particular, for $^{73}$Ge this condition is met for $q$ $\leq29+$, for which $100$ K-capture channels have been unveiled.

\clearpage
\section{NEEC in laser-plasma scenario with GSA}

Following the theory presented in Ref. \onlinecite{Gunst2018}, it is possible to compare the reaction rates provided by NEEC and the process of  direct photoexcitation.
In particular, considering the parameters reported in Table \ref{tab:laser_par} and using the first scaling law of Ref. \onlinecite{Gibbon1996}, we obtain $T_\mathrm{e}=\SI{2.5}{\kilo\electronvolt}$ and $n_\mathrm{e}=\SI{9e19}{\per\cubic\centi\meter}$ as plasma temperature and electron density, respectively.

\begin{figure*}[!h]
\includegraphics[width=\linewidth]{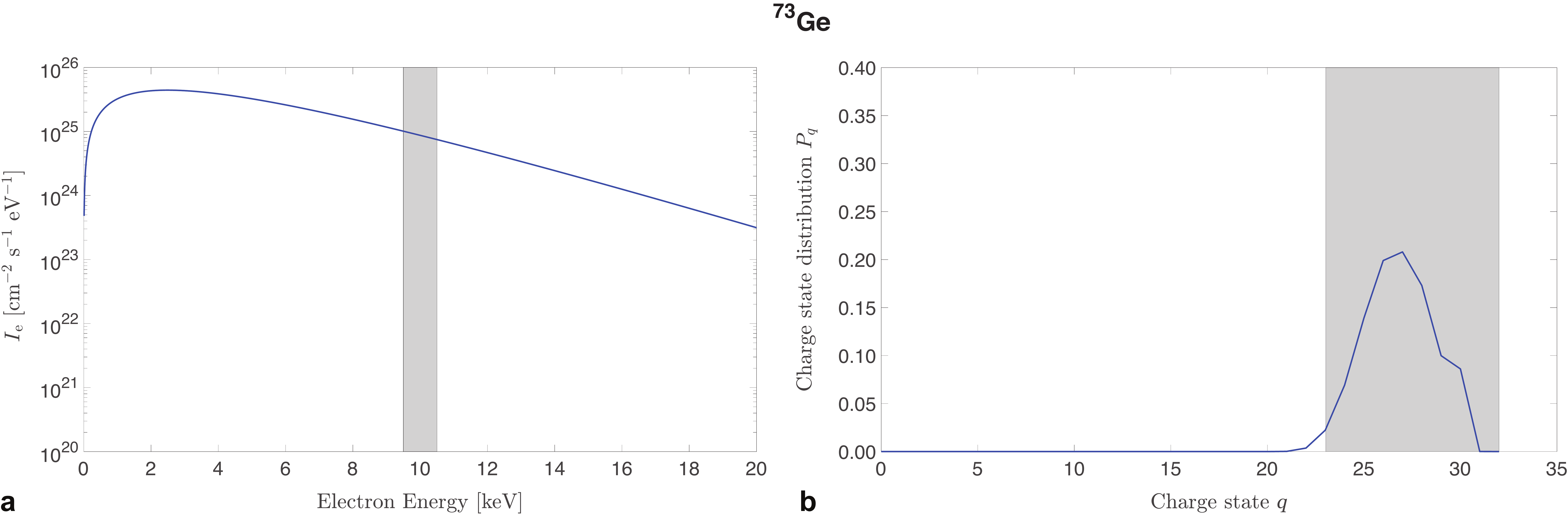}
\caption{\label{fig:Plasma_scenario} (a) Relativistic electron flux distribution produced by a \SI{2.5}{\kilo\electronvolt} plasma temperature. Gray box highlights the L-channels energy range. (b) Ion charge state distribution obtained with FLYCHK. The gray box indicates the charge state interval $q= \interval{23+}{32+}$, for which the resonance strengths of Table \ref{tab:Ge73} have been evaluated.}
\end{figure*}

\begin{table*}[!ht]
\caption{\label{tab:laser_par} Laser characteristics and absorption coefficient.}
\begin{ruledtabular}
\begin{tabular}{ll}
{$E_{\textrm{pulse}}$} & {$\SI{0.7}{\milli\joule}$}\\
{$R_{\textrm{focal}}$} & {$\SI{6.5}{\micro\meter}$}\\
{$\tau_{\textrm{pulse}}$}&{\SI{50}{\femto\second}}\\
{$\lambda$} & {$\SI{800}{\nano\meter}$}\\
{$f$} & {0.1}\\
\end{tabular}
\end{ruledtabular}
\end{table*}
\noindent
By means of the radiative collisional code FLYCHK \cite{Chung2005} it is possible to determine the ion charge state distribution, considering a non-local thermodynamical equilibrium steady state for the plasma. Fig. \ref{fig:Plasma_scenario} shows the results obtained  for the electron flux $I_\mathrm{e}$ and the charge state distribution $P_q$ considering this scenario.

\noindent
Continuing the derivation of Eq. \ref{eq:ResStrength}, the partial NEEC rate for the capture level $\alpha_\mathrm{r}$ in an ion characterized by a charge state $q$ can be written as \cite{Gunst2018}:

\begin{equation}
\lambda_{\mathrm{NEEC}}^{q,\alpha_\mathrm{r}}= S_{\mathrm{NEEC}}^{q,\alpha_\mathrm{r}}\ I_\mathrm{e}(E_{\alpha_\mathrm{r}})\ ,
\end{equation}
\noindent
where $E_{\alpha_\mathrm{r}}$ represents the resonance energy of the capture channel $\alpha_\mathrm{r}$ and $I_\mathrm{e}$ the electron flux.
Considering NEEC into ions which are in their ground state (GSA), the total NEEC rate can be written as summation over all the capture channels $\alpha_\mathrm{r}$ and all over the charge state $q$ present in the plasma:
\begin{equation}
\lambda_{\mathrm{NEEC}}= \sum_q \sum_{\alpha_\mathrm{r}} P_q\  \lambda_{\mathrm{NEEC}}^{q,\alpha_\mathrm{r}}\ .
\label{eq:NEEC_Rate}
\end{equation}
\noindent
As evidenced in Fig. \ref{fig:Plasma_scenario}b, more than $99.6\%$ of the ions fall in the range $q=\interval{23+}{32+}$, thus a good estimate of the total NEEC rate, in case of GSA, can be obtained using the resonance strengths reported in Table \ref{tab:Ge73}. In the presented experimental scenario, this results in a total NEEC rate $\lambda_{\mathrm{NEEC}}=\SI{2.63e-3}{\per\second}$, while for direct photoexcitation the total rate is $\lambda_{\gamma}=\SI{6.65e-1}{\per\second}$. The strongest resonance channel of Table \ref{tab:Ge73} is contributing to the partial NEEC rate as  $\lambda_{\mathrm{NEEC}}^{32+,2\mathrm{p}_{3/2}}=\SI{3.03e-3}{\per\second}$ and $P_q = \SI{2.97e-8}{}$. Even if $P_q$ of this channel would have been 1, it would have been not enough to compensate the photoexcitation rate $\lambda_{\gamma}$.

In case of NEEC-EXI, it is necessary to sum also all over the initial electronic configurations $\alpha_\mathrm{0}$, thus Eq. \ref{eq:NEEC_Rate} modifies as following:

\begin{equation}
\lambda_{\mathrm{NEEC}}= \sum_q \sum_{\alpha_\mathrm{0}} \sum_{\alpha_\mathrm{r}} P_q^{\alpha_\mathrm{0}}\ \lambda_{\mathrm{NEEC}}^{q,\alpha_\mathrm{0},\alpha_\mathrm{r}} .
\label{eq:NEEC_Rate_EXI}
\end{equation}
\noindent

A direct comparison with NEEC-EXI rate is not straightforward and beyond the scope of this paper. Indeed, a dynamical study of the plasma formation and expansion is needed through particle-in-cell (PIC) codes. Nonetheless, it is worth to mention that for the resonance energy of the K-channels the electron flux provided by the plasma, evaluated in Fig. \ref{fig:Plasma_scenario}a, is  few orders of magnitude less than that for the L-channels range. In this case, either a more proper choice of the electron temperature has to be done, or a free electron source in resonance with the desired channels has to be provided. For example, considering the $1\mathrm{s}^1\ 2\mathrm{s}^1\ 3\mathrm{p}_{3/2}^1$ electronic configuration at $q=29+$, assuming $P_q^{\alpha_0} \simeq 1$ and an electron flux of \SI{1e25}{\per\square\centi\meter\per\second\per\electronvolt} at low energy (e.g., with an optimized external electron source), the total NEEC rate could reach the value of $\lambda_{\mathrm{NEEC}}=\SI{8.64}{\per\second}$ for the highest resonance strength of the K-shell, significantly higher than the one obtained with the direct photoexcitation. In these circumstances, the external source can act as an \textit{electron switch} that boosts the isomer depletion.


\input{Supplemental.bbl}

%% file: Manuscript.bbl
%

%% file: Supplemental.bbl
%